# CuBr$_2$ － A New Multiferroic Material with High Critical Temperature

*L. Zhao$^1$, T. L. Hung$^1$, C. C. Li$^1$, Y.Y. Chen$^1$, M. K. Wu$^{1*}$, R.K. Kremer$^2$, M. G. Banks$^2$, A. Simon$^2$, M.H. Whangbo$^3$, C. Lee$^3$, J. S. Kim$^4$, I. Kim$^5$,* and *K.H. Kim$^5$*

[1] Prof. M.K. Wu* Corresponding-Author, Dr. L. Zhao, T.L. Hung, C.C Li, Prof. Y.Y. Chen
Institute of Physics, Academia Sinica,
Taipei 11529, Taiwan
E-mail: mkwu@phys.sinica.edu.tw

[2] Dr. R.K. Kremer, Dr. M.G. Banks, Prof. A. Simon,
Max Planck Institute for Solid State Research, 70569 Stuttgart, Germany

[3] Prof. M.-H. Whangbo, Dr. C. Lee,
Department of Chemistry, North Carolina State University, Raleigh, North Carolina 27695-8204, U.S.A.

[4] Prof. J. S. Kim
Department of Physics, Pohang University of Science and Technology, Pohang 790-784, Korea

[5] I. Kim, Prof. Kee Hoon Kim
CeNSCMR, Department of Physics and Astronomy, Seoul National University, Seoul 151-747, Korea

Recently, multiferroic materials, with magnetic and electric polarization coexisting and being strongly coupled to each other, have attracted unprecedented attention[1,2] especially since the spin-driven ferroelectricity and concomitant sizeable magnetoelectric coupling were discovered in some magnetically frustrated manganites.[3, 4] Research into multiferroics opens new opportunities not only for fundamental physics of strongly correlated materials, but also for    potential application of highly efficient multi-functional devices as sensors and memory devices.[5, 6] In these multiferroic materials, the magnetoelectric coupling goes along with complex spin structures, which are usually non-collinear and break the inversion symmetry of the lattice. The spontaneous electric polarization emerges concurrently, and strong magnetoelectric coupling effect can be observed.[2] For example, the electric polarization can be reversed or rotated by external magnetic field and vice versa, which is crucial for developing future memory devices.

A well-accepted key microscopic mechanism for multiferroicity comes down to the inverse Dzyaloshinskii-Moriya(DM) interaction, which is an antisymmetric relativistic correction to the superexchange coupling.[7] It can also be expressed in an equivalent spin current picture by Katsura, Nagaosa and Balatsky, (KNB model).[8] The microscopic polarization $P_{ij}$ induced by neighboring spins can be formulated as $P_{ij}=A\hat{e}_{ij}\times(S_i\times S_j)$, where the coupling coefficient $A$ is determined by the spin-orbit coupling and exchange interactions, and $\hat{e}_{ij}$ is the unit vector connecting the sites $i$ and $j$. This expression is consistent with the phenomenological theory based on symmetry consideration,[9] and has worked well for, e.g., the helimagnets such as TbMnO$_3$.[10] The sign and magnitude of $A$ for real systems can be calculated in the generalized KNB model.[11]

Besides manganites, spin-driven ferroelectricity has also been found in some other helimagnetic systems, mostly focused on oxides, including cuprates,[12-14] ferrates, [15] nickelates.[16] Since magnetic frustration due to highly competing magnetic interactions usually reduces the critical temperature of spin-ordered phases, the transition temperature of most known multiferroic materials is low (usually below 40 K), and strong magnetoelectric coupling usually exists only in a narrow temperature range. Until now, only a few compounds have been found with high critical temperatures. A famous example is copper(II) oxide, CuO. The ferroelectricity is tied to its incommensurate spiral magnetic structure (AF2 phase) existing between 213 and 230 K.[12] In some hexaferrites, multiferroicity even above room temperature has been reported.[17] But the high dielectric loss (tan δ >> 1) due to semiconducting electronic properties prevents accurate dielectric measurements and future practical applications. Until now, the reproducibility of single-phase room temperature multiferroics still seems as elusive as room-temperature superconductors.[6] At present, there is still a long way to go in developing applicable multiferroic materials with both high critical temperature and low electric loss.

So far, non-oxide based magnetic systems have been rarely studied in the research of multiferroic materials. In 2009, Banks *et al.* investigated the magnetic structure of copper(II) chloride, $CuCl_2$.[18] Anhydrous $CuCl_2$ is a chemically simple quasi-1D antiferromagnetic spin 1/2 quantum chain system, which consists of parallel $CuCl_2$ ribbons made up of $CuCl_4$ squares via edge-sharing which resembles the $CuO_4$-square-based chain multiferroic cuprates as $LiCu_2O_2$ and $LiCuVO_4$.[13,14] Neutron diffraction experiments revealed an incommensurate spin spiral structure propagating along the ribbons below its Néel temperature $T_N$ ~24 K. Banks *et al.* predicted possible multiferroicity existing in this non-oxide spiral magnet. Later, the measurement by Seki *et al.* confirmed the spin-driven ferroelectricity and strong magnetoelectric coupling in $CuCl_2$.[19] $CuCl_2$ becomes the first member of halogenide-based multiferroic family. Although its critical temperature is still low, it motivates further search for higher temperature multiferroics. In the following, we report the first observation of multiferroicity in anhydrous copper(II) bromide, $CuBr_2$, with a high transition temperature close to liquid nitrogen temperature ($T_N$ = 73.5K). Large magnetoelectric coupling and low dielectric loss are found to coexist in $CuBr_2$ below $T_N$.

Like $CuCl_2$, $CuBr_2$ crystallizes in a $CdI_2$-type structure with a monoclinic lattice of space group $C2/m$ ($a$ = 7.21 Å, $b$ = 3.47 Å, $c$ = 7.05 Å, $\beta$ = 119.6°).[20] As shown in Fig. 1(a), layers of $CuBr_2$ stack along the *c*-axis direction. In each $CuBr_2$ layer, ribbons made up of edge-sharing $CuBr_4$ squares run along *b*-axis. The bonding angle $Cu^{2+}$-$Br^-$-$Cu^{2+}$ along the $CuBr_2$ ribbon is close to 90°. The nearest-neighbor exchange is expected to be ferromagnetic according to the Anderson-Kanamori-Goodenough rules, and the next-nearest-neighbor exchange is antiferromagnetic according to the rule for super-superexchange interaction.[21] The magnetic properties of $CuBr_2$ remained largely unknown until now. Barraclough *et al.* measured the magnetic susceptibility of $CuBr_2$ in 1964 and observed a broad maximum at 226 K, but no clear evidence for a long-range magnetic ordering was reported.[22] Bastow *et al.* determined the specific heat and found an anomaly at ~74 K which they ascribed to the onset of a magnetic ordering.[23]

In analogy to the properties of isostructural $CuCl_2$ and similar $CuO_2$-ribbons systems like $LiCu_2O_2$ and $LiCuVO_4$, [13,14] we can reasonably well expect a quasi-1D spiral spin ground state with a long-range incommensurate order in $CuBr_2$, arising from the competing magnetic exchange interactions. In the following we demonstrate that this is in fact the case and multiferroicity is found in $CuBr_2$ below its Néel temperature of $T_N = 73.5$ K.

Fig. 1(c) shows the magnetic susceptibility ($\chi$) of $CuBr_2$. A broad hump appears in $\chi(T)$ with a maximum around ~200 K. As typically found in low-dimensional spin systems, an extended antiferromagnetic short-range ordering leads to this susceptibility hump high above $T_N$. The onset of a long-range magnetic ordering is indicated by a kink-like anomaly at $T_N$. At lower temperature (below 20 K), $\chi(T)$ up-turns sharply. The ZFC and FC curves merge, excluding possible ferromagnetism. For clarity, the corresponding temperature derivative, $d\chi/dT$, is shown in the inset. A $\lambda$-shaped peak is clearly revealed at $T_N$. Further confirmation of the long-range ordering comes from the specific heat ($C_p$) measurement, which shows a similar $\lambda$-shaped anomaly at 73.5 K (see Fig. 3(d)). The change of entropy estimated from the anomalous peak in $C_p(T)$ (see the inset) amounts to ~0.05 J K$^{-1}$ mol$^{-1}$, much less than the full magnetic entropy $R \ln 2$ (~5.76 J K$^{-1}$ mol$^{-1}$) as expected in a $S =$

1/2 spin system, indicating the persistence of strong short-range ordering high above $T_N$.

A comparison of the neutron powder diffraction patterns collected on a polycrystalline sample at 1.8 K and 80 K reveals weak magnetic Bragg reflections with intensity by two orders of magnitude smaller than the nuclear reflections (Fig. 1(b)). They disappear as $T > T_N$. Analogous to $CuCl_2$ the magnetic reflections can be indexed based on a C-centered monoclinic cell with a propagation vector $\tau = (1, 0.2350(3), 0.5)$, indicating an incommensurate magnetic structure with an approximate quadrupling of the nuclear cell along b and a pitch angle of ~85° between nearest-neighbor Cu moments. The refinement of the magnetic structure was successfully performed using the difference of the patterns collected at 1.8 and 80 K with the Cu atoms in the Wyckoff position (2a) (0,0,0) and assuming a helical spin spiral propagating along the b-axis. In the magnetic refinement the overall scale factor was fixed to the value obtained from a refinement of the 80 K nuclear structure. Within error bars the components of the refined magnetic moment is consistent with a circular spiral (magnitude of the moment 0.39(3) $\mu_B$) rotating within the plane of the $CuBr_2$ ribbons, similar to what has been found for $CuCl_2$.[18]

Fig. 2(a) summarizes the temperature-dependent dielectric constant, $\varepsilon(T)$, measured in zero field and at different testing frequencies (1 kHz to 1 MHz). The most remarkable feature is the sharp rising in $\varepsilon(T)$ just below $T_N$, which coincides well with the anomalies observed in the magnetization and heat capacity measurements

(see Fig. 1). With decreasing temperature, $\varepsilon(T)$ rises quickly and passes through a broad maxima (around 55 K), followed by a slow decrease towards low temperatures. We have confirmed this behavior by repeating the temperature cycles and on samples of different batches. Therefore possible extrinsic factors such as trapped interfacial charge carriers and thermal degradation of bulk sample can be excluded, indicating an intrinsic electric transition of $CuBr_2$ at $T_N$. For convenience, we only adopt the data measured at 100 kHz in our following discussions.

The temperature-dependent dielectric loss (tan$\delta$) (Fig. 2(b)) also shows a similar anomaly at $T_N$, independent of testing frequencies as well. It is noticeable that the dielectric loss is quite small (<<0.1) due to the highly insulating nature of the $CuBr_2$ sample. The concurrence of the anomalies in the dielectric constant and corresponding loss suggest possible ferroelectricity below $T_N$, which is confirmed in our later pyroelectric measurements.

A typical result of a pyroelectric current measurement is shown in Fig.2(c). A clear $\lambda$-shaped pyroelectric current peak emerges as the temperature approaches $T_N$. The current drops sharply to zero as $T > T_N$. The electric polarization acquired by integrating the pyroelectric current is switched with the opposite poling electric field (not shown), proving its ferroelectric nature as $T < T_N$. At $T = 10$ K, the saturated polarization reaches 8.0 μC m$^{-2}$, which is of the same order as in $LiCu_2O_2$ or $CuCl_2$.[13, 19]

We further investigated the magnetoelectric coupling effect in $CuBr_2$ in an external magnetic field up to 9T, which is parallel to the plate-like sample (i.e., $H \perp E$ configuration). We also tested the $H \parallel E$ measuring configuration in our experiments. No notable difference was observed, since intrinsic anisotropy of $CuBr_2$ is canceled out in the polycrystalline samples. Fig.3 (a) shows $\varepsilon(T)$ measured in different $H$ (0-9 T). The broad dielectric hump is gradually suppressed by increasing $H$, while the transition temperature remains almost unchanged.

To quantify this suppressing effect by $H$ in $CuBr_2$, we measured the magnetodielectric coefficient (*MC*), defined as $\Delta\varepsilon(H)/\varepsilon(H=0)$ ($\Delta\varepsilon(H) = \varepsilon(H) - \varepsilon(H=0)$), at several temperatures above and below $T_N$, as shown in Fig. 3(b). Above $T_N$ (e.g., 90K), the *MC* is very close to zero. A significant magnetodielectric effect appears only below $T_N$. As $T < T_N$, the *MC* is negative and its magnitude increases monotonously as $H$ grows, and the *MC* never saturates till $H = 9$ T. Close to the maxima of the dielectric anomaly, the magnetodielectric effect is the most remarkable. At $T = 55$ K, the *MC* is -23% for $H = 9$ T. The *MC* decays slightly at lower temperatures. At the lowest temperature ($T = 10$ K), the *MC* still reaches −15 % at $H = 9$ T, which is quite high compared with most known multiferroic materials, including $CuCl_2$ and $LiCu_2O_2$. Although in some rare-earth manganites[3,4] and ferroborates[24], even higher *MC*s have been reported, they mostly appear in a narrow temperature range close to the magnetic transition or show a non-monotonic field-dependence.

Although $H$ considerably suppresses the anomaly in dielectric permittivity in the multiferroic state ($T < T_N$), it is surprising that $H$ enhances the ferroelectric polarization greatly as shown in Fig. 3(c). At $T = 10$ K, $P$ increases from 8.0 (in $H=0$) to 22.5 Cm$^{-2}$ (in $H = 9$ T), although the transition temperature remains almost unchanged. Such a giant enhancement was also reported recently for LiCu$_2$O$_2$ single crystals with $H$ along certain directions,[25] due possibly to the modulation or distortion of the spin helix by $H$. Considering the polycrystalline character of our sample, it is possible that the intrinsic magnetoelectric coupling as $H$ along certain directions is expected much stronger than the above results from bulk samples.

The observed concurrence of ferroelectric transition and magnetic ordering reveals that CuBr$_2$ constitutes a new multiferroic material. Compared with most known multiferroic materials, CuBr$_2$ exhibits a very high critical temperature ($T_N = $ 73.5 K, close to the boiling point of liquid nitrogen). Multiferroicity in CuBr$_2$ spans a very broad temperature range (0 K $< T < T_N$). Additionally, the low dielectric loss makes anhydrous CuBr$_2$ a very promising material for developing new magnetoelectric devices. Our finding will stimulate related research on the anion effect in these low-dimensional spin systems and further improve our understanding of the physics of multiferroicity.

In summary, we have demonstrated that the spiral spin ordering below $T_N = $ 73.5 K leads to multiferroicity in CuBr$_2$. The high critical temperature, low dielectric loss and the strong magnetoelectric coupling make $CuBr_2$ a very promising multiferroic material. Our studies emphasize the importance of the anion effect in these low-dimensional spin systems, and suggest routes to stimulate the further search for high temperature multiferroic materials.

*Experimental*

Anhydrous $CuBr_2$ powder of high purity is commercially available. In our experiments no difference was observed on the samples from different suppliers (Alfa Aesar, Acros, and Sigma-Aldrich). Because of its hygroscopic nature, the samples were handled in a Ar-filled glove box most of the time.

Magnetic properties were determined with a SQUID magnetometer (MPMS-XL). The crystalline flakes was use to measure the heat capacity on a PPMS (Physical Properties Measurement System, Quantum Design, Inc.). Neutron diffraction patterns were collected at ILL's high-intensity medium resolution powder diffractometer D20 at Grenoble. Refinement of the powder diffraction patterns was performed using the FULLPROF suite.

For the dielectric measurement, powder of about 0.5g is hot-pressed (10 ton, ~150 °C, ~1 h) into a dense pellet with a diameter of 10 mm, which subsequently is cut and polished to a thin rectangular plate with thickness of about 0.3 mm. We use silver paint attached to both sides as electrodes to form a parallel plate capacitor and measure its capacitance with a high-precision capacitance meter. The main sources of error such as residual impedances in the whole circuit are carefully compensated.

The corresponding electric polarization is obtained from the integration of the measured pyroelectric current. While cooling the sample is polarized with an applied electric field of about 500 kV m$^{-1}$. At the base temperature, we then remove the electric field and short-circuit the sample for one hour to remove possible stray charges. The pyroelectric current is measured during the warming process at different rates (1-4 K min$^{-1}$).

*Acknowledgements*

LZ acknowledges the financial support from the National Science Council of Taiwan. Work at SNU was supported by national CRI program (2010-0018300). This work was also supported by the Leading Foreign Research Institute Recruitment program (2010-00471) through NRF funded by MEST.

**Figure 1.** (a) Crystal structure of $CuBr_2$. (b) Neutron powder diffraction patterns of $CuBr_2$. Main frame: difference between the 2 K and 80 K patterns. The top panel shows a Cu spin spiral chain along the $CuBr_2$ ribbon (b-axis). The inset displays the powder diffraction pattern ($\lambda = 2.41$ Å) at 2 K together with a FULLPROF profile refinement (solid line) with the Bragg angles of the nuclear and the magnetic Bragg peaks indicated by the vertical bars indicated below the pattern. (c) Magnetic susceptibility of $CuBr_2$. The inset shows the corresponding temperature derivative. (d) Specific heat of $CuBr_2$ with the anomaly at $T_N$.

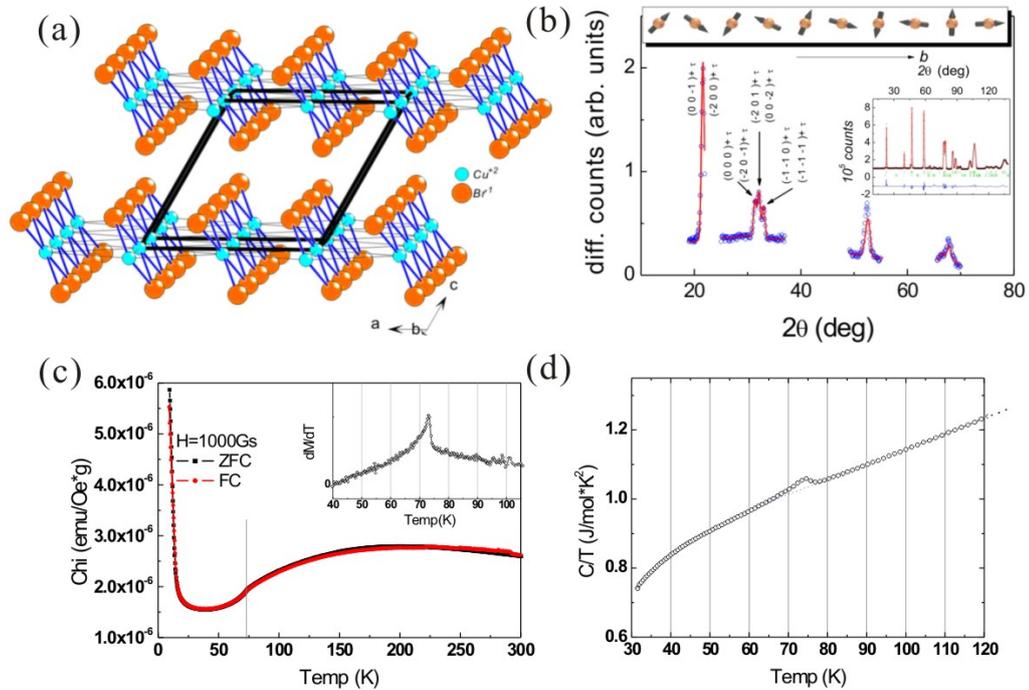

**Figure 2.** (a) $\varepsilon(T)$ of a typical $CuBr_2$ sample measured at different frequencies in $H = 0$. (b) $\varepsilon(T)$ and the corresponding tan loss measured at frequency = 100 kHz in $H = 0$. (c) Pyroelectric current measured with a warming rate of 3 K min$^{-1}$ under $H = 0$. The corresponding polarization is acquired by integrating the pyroelectric current.

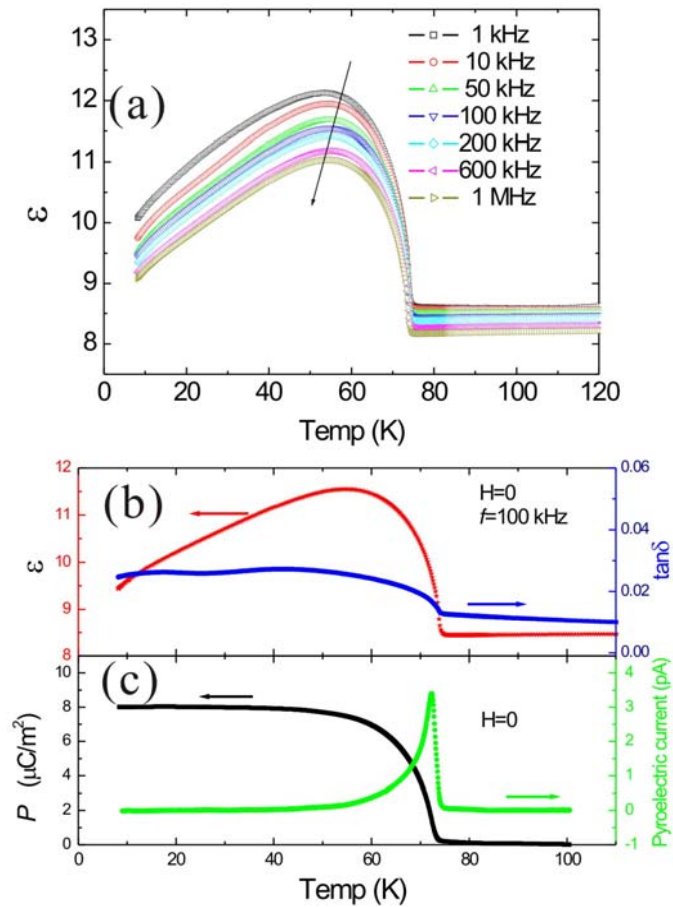

**Figure 3.** (a) Temperature dependent dielectric constant and (c) electric polarization of $CuBr_2$ measured in different $H$ (0 - 9 T). (b) Magnetodielectric coefficients of $CuBr_2$ at different temperatures.

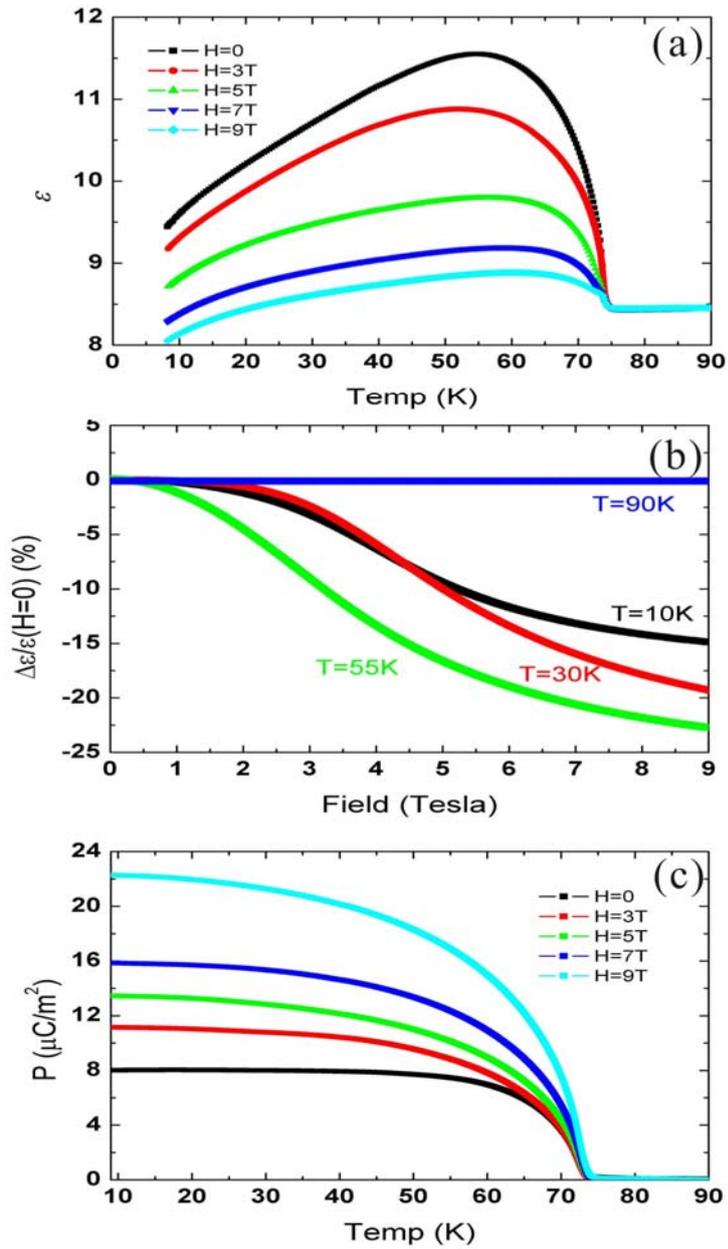